\documentclass[11pt]{article}
\usepackage{amssymb}
\usepackage{amsmath}
\usepackage[english]{babel}
\usepackage{graphicx}
\usepackage{cite}

\def\Journal#1#2#3#4{{#1} {\bf #2}, #3 (#4)}

\def\NCB{\em Nuovo Cim. B}

\def\PLB{{\em Phys. Lett.}  B}
\def\PRL{\em Phys. Rev. Lett.}
\def\PRD{{\em Phys. Rev.} D}

\def\GaC{\em Gravitation and Cosmology}
\def\GaCS{{\em Gravitation and Cosmology} Supplement}

\def\JETPL{\em JETP Lett.}
\def\PAN{\em Phys.Atom.Nucl.}
\def\CQG{\em Class. Quantum Grav.}

\def\MPLA{{\em Mod. Phys. Lett.}  A}
\def\IJTP{\em Int. J. Theor. Phys.}
\def\NJP{\em New J. of Phys.}

\def\JCAP{\em JCAP}

\def\BWP{\em Bled Workshops in Physics}
\def\JPCS{{\em J. Phys.:} Conf. Ser.}

\def\IJMPA{{\em Int. J. Mod. Phys.}  A}
\def\IJMPD{{\em Int. J. Mod. Phys.}  D}

\def\AHEP{\em Adv. High Energy Phys.}
\def\s{{\,\rm s}}

\def\({\left(}
\def\){\right)}

\def\beq{\begin{equation}}
\def\eeq{\end{equation}}
\def\bea{\begin{eqnarray}}
\def\eea{\end{eqnarray}}

\begin{document}

    \begin{center}
        \large \textbf{Decaying Dark Atom constituents and cosmic positron excess}
    \end{center}

    \begin{center}
   K.Belotsky$^{1,2}$, M. Khlopov$^{1,2,3}$, C. Kouvaris$^{4}$, M.Laletin$^{1}$

    \emph{$^{1}$National Research Nuclear University ``Moscow Engineering Physics Institute", 115409 Moscow, Russia \\
    $^{2}$ Centre for Cosmoparticle Physics ``Cosmion" 115409 Moscow, Russia \\
$^{3}$ APC laboratory 10, rue Alice Domon et L\'eonie Duquet \\75205
Paris Cedex 13, France\\
$^4$$\text{CP}^3$-Origins, University of Southern Denmark, Campusvej 55, Odense 5230, Denmark}

    \end{center}

\medskip

\begin{abstract}

We present a scenario where dark matter is in the form of dark atoms that can accomodate the experimentally observed excess of positrons in PAMELA and AMS-02 while being compatible with the constraints imposed on the gamma-ray flux from Fermi/LAT. This scenario assumes that the dominant component of dark matter is in the form of a bound state between a helium nucleus and a $-2$ particle and a small component is in the form of a WIMP-like dark atom compatible with direct searches in underground detectors. One of the constituents of this WIMP-like state is a $+2$ metastable particle with a mass of 1 TeV or slightly below that by decaying to $e^+e^+$, $\mu^+ \mu^+$ and $\tau^+ \tau^+$ produces the observed positron excess. These decays can naturally take place via GUT interactions. If it exists, such a metastable particle can be found in the next run of LHC. The model predicts also the ratio of leptons over baryons in the Universe to be close to -3. \\[.1cm] {\footnotesize \it Preprint: CP$^3$-Origins-2014-005
  DNRF90 \& DIAS-2014-5.}

\end{abstract}
\section{Introduction}
The possibility of dark matter being in the form of ``dark atoms'' has been studied extensively~\cite{Blin1,Blin2,shadow2,hodges,GH,BM,BDM,FV,MT,foot,SZ,arkani,kaplan,behbahani,kaplan2,cline,cline2,cyr,cyr2,dweinberg,cline3}. In this scenario  new stable particles are bound by new dark forces (like mirror partners of ordinary particles bound by mirror electromagnetism~\cite{LeeYang,KOP,ZKrev,OkunRev,Paolo}). However, it turns out that even stable electrically charged particles can exist hidden in dark atoms,  bound by  ordinary Coulomb interactions (see \cite{Levels,Levels1,mpla,DMRev} and references therein).
Stable particles with charge -1 (and
corresponding antiparticles as tera-particles \cite{Glashow}) are excluded due to overproduction of anomalous isotopes.  However, negatively doubly charged particles  are not constrained by anomalous isotope searches as much as -1 charged particles~\cite{Fargion:2005xz}.
There
exist several types of particle models where heavy
stable -2  charged species, $O^{--}$, are predicted:
\begin{itemize}
\item[(a)] AC-leptons, predicted
as an extension of the Standard Model, based on the approach
of almost-commutative geometry \cite{Khlopov:2006dk,5,FKS,bookAC}.
\item[(b)] Technileptons and
anti-technibaryons in the framework of Walking Technicolor
(WTC) \cite{KK,Sannino:2004qp,Hong:2004td,Dietrich:2005jn,Dietrich:2005wk,Gudnason:2006ug,Gudnason:2006yj}.
%\item[(c)] and, finally, stable "heavy quark clusters" $\bar U \bar U \bar U$ formed by anti-$U$ quark of 4th generation
% \cite{Khlopov:2006dk,Q,I,lom,KPS06,Belotsky:2008se}.
\end{itemize}
All these models also
predict corresponding +2 charge particles. If these positively charged particles remain free in the early Universe,
they can recombine with ordinary electrons in anomalous helium, which is strongly constrained in
terrestrial matter. Therefore a cosmological scenario should provide a  mechanism which suppresses anomalous helium.
There are  two possible mechanisms than can provide a suppression:
\begin{itemize}
\item[(i)] The abundance of anomalous helium in the Galaxy may be significant, but in terrestrial matter
 a recombination mechanism could suppress this abundance below experimental upper limits \cite{Khlopov:2006dk,FKS}.
The existence of a new U(1) gauge symmetry, causing new Coulomb-like long range interactions between charged dark matter particles, is crucial for this mechanism. This leads inevitably to the existence of dark radiation in the form of hidden photons.
\item[(ii)] Free positively charged particles are already suppressed in the early Universe and the abundance
of anomalous helium in the Galaxy is negligible \cite{mpla,I}.
\end{itemize}
These two possibilities correspond to two different cosmological scenarios of dark atoms. The first one is
realized in the scenario with AC leptons, forming neutral AC atoms \cite{FKS}.
The second assumes a charge asymmety  of the $O^{--}$ which form the atom-like states with
primordial helium \cite{mpla,I}.

If new stable species belong to non-trivial representations of
the SU(2) electroweak group, sphaleron transitions at high temperatures
can provide the relation between baryon asymmetry and excess of
-2 charge stable species, as it was demonstrated in the case of WTC
\cite{KK,KK2,unesco,iwara}.

 After formation
in the  Big Bang Nucleosynthesis (BBN), $^4He$ screens the
$O^{--}$ charged particles in composite $(^4He^{++}O^{--})$ {\it
$OHe$} ``atoms'' \cite{I}.
In all the models of $OHe$, $O^{--}$ behaves either as a lepton or
as a specific ``heavy quark cluster" with strongly suppressed hadronic
interactions. Therefore $OHe$ interactions with matter are
determined by the nuclear interactions of $He$. These neutral primordial
nuclear interacting objects can explain the modern dark matter
density and represent a nontrivial form of strongly
interacting dark matter \cite{McGuire:2001qj,McGuire1,McGuire2,Starkman,Starkman2,Starkman3,Starkman4,Starkman5,Starkman6}.

The cosmological scenario of the $OHe$ Universe can explain many results of experimental searches for dark matter \cite{mpla}. Such a scenario is insensitive to the properties of $O^{--}$, since the main features of the $OHe$ dark atoms are determined by their nuclear interacting helium shell. In terrestrial matter such dark matter species are slowed down and cannot cause significant nuclear recoil in the underground detectors, making them elusive in direct WIMP search experiments (where detection is based on nuclear recoil) such as CDMS, XENON100 and LUX~\cite{CDMS,CDMS2,CDMS3,xenon,lux}. The positive results of DAMA and possibly CRESST and CoGeNT experiments \cite{DAMA,DAMA-review,Bernabei:2008yi,cresst,cogent} can find in this scenario a nontrivial explanation due to a low energy radiative capture of $OHe$ by intermediate mass nuclei~\cite{mpla,DMRev}.

It has been also shown \cite{KK,KK2,unesco,iwara} that a two-component dark atom scenario is also possible. Along with  the dominant $O^{--}$ abundance, a much smaller excess of positively doubly charged techniparticles can be created. These positively charged particles are hidden in WIMP-like atoms, being bound to $O^{--}$. In the framework of WTC such positively charged techniparticles can be metastable, with a dominant decay channel to a pair of positively charged leptons. In this paper we show that even a $10^{-6}$ fraction of such positively charged techniparticles with a mass of 1 TeV or less and a lifetime of~$10^{20} \s$,  decaying   to $e^+e^+$, $\mu^+ \mu^+$, and $\tau^+ \tau^+$  can explain the observed excess of cosmic ray positrons, being compatible with the observed gamma ray background.

One should note that as it was shown in \cite{FKS,KK,I,KK2}  (for a review see \cite{mpla,Khlopov:2006dk} and  references therein) the case of -2 charged stable particles is significantly different from the case of stable or metastable particles with charge -1, avoiding severe constraints on charged particles from anomalous isotope searches and BBN due to their catalytic effects  (see e.g. \cite{ATPTnew,CLLCMPnew,BBNPNPnew}). In the essence this difference comes from the fact that primordial He formed in BBN, captures -2 charged particles in neutral OHe states, while -1 charged particles are captured by He in +1 charged ions, which either (if stable) form anomalous isotopes of hydrogen, or (if long-lived, but metastable) catalyze processes of light element production and influence their abundance. Nuclear physics of OHe is in the course of development, but a qualitative analysis has shown \cite{unesco} that the OHe interactions with matter should not lead to overproduction of anomalous isotopes, while OHe catalytic effects in BBN can lead to primordial heavy element production, but not to overproduction of light elements.

The paper is organized as follows: In section 2 we give a brief review of dark atoms made of stable charged techniparticles. In Section 3 we present the constraints and the predictions of the scenario with respect to the parameters of the Technicolor model we use as well as how the  ratio of lepton over baryon number is deduced. In section 4 we show what GUT operators can implement the decay of the doubly charged particle to leptons. In section 5, we show how the scenario of decaying dark matter can be realized, and how it can explain the PAMELA and AMS-02 results while satisfying the Fermi/LAT constraints. We conclude in section 6.

\section{Dark atoms from Techniparticles}

Technicolor theories that do not violate the electroweak precision tests, while not introducing large flavor changing currents, have been extensively studied lately~(see~\cite{Sannino:2009za} and references therein).~Old models where fermions transformed under the fundamental representation of the gauge group, required a large number of flavors (for a given number of colors) in order to be close to the conformal window and thus to suppress the flavor changing neutral currents. The need for many flavors coupled to the electroweak sector (that violates the electroweak precision measurements) disfavored Technicolor in the past. However it has been demonstrated that once one allows fermions to transform under higher representations of the gauge group, quasi-conformality can be achieved even with a small number of colors and flavors~\cite{Sannino:2004qp,Hong:2004td,Dietrich:2005jn}. This means that there is a set of Technicolor models that evade the strict constraints of the electroweak tests, making Technicolor a viable candidate for the TeV energy scale. Apart from the perturbative calculation of the oblique parameters~\cite{Dietrich:2005wk} in this type of models, non-perturbative calculations based on holographic descriptions~\cite{Dietrich:2008ni,Dietrich:2008up,Dietrich:2009af} showed that indeed the oblique S parameter can be small. Note that the oblique parameters (e.g. S, T, and U) measure the modifications of the Standard Model gauge boson vacuum polarization amplitudes caused by contributions of new physics.~These parameters are severely constrained by electroweak precision tests. Extra flavors that couple to the electroweak sector contribute to these parameters and can potentially exclude a model.

One of the simplest models that possesses the features described above, is the so called Minimal Walking Technicolor~\cite{Sannino:2004qp,Gudnason:2006ug,Foadi:2007ue}. The theory consists of two techniquarks transforming under the adjoint representation of an $SU(2)$ gauge group, and an extra family of leptons $\nu'$ and $\zeta$ coupled to the electroweak in order to cancel the global Witten anomaly. The hypercharge assignment can be chosen consistently (without introducing gauge anomalies) such that one of the techniquarks has zero electric charge. Such a simple theory can have a variety of dark matter candidates, ranging from dark matter particles that are Goldstone bosons of the theory (with nonzero technibaryon number)~\cite{Gudnason:2006yj,Ryttov:2008xe,Frandsen:2009mi}, or Majorana WIMPs~\cite{Kouvaris:2007iq,Kainulainen:2006wq,Kouvaris:2008hc,Belotsky:2008vh,Antipin:2009ks,Hapola:2013mba}. Apart from these possibilities, there is another intriguing scenario, that of an electromagnetic bound state between a $+2$ charged helium nucleus with a $-2$ charged techniparticle~\cite{KK,KK2}. More specifically in~\cite{KK}, we examined the possibility where the dark matter bound state is $He\bar{U}\bar{U}$ or $He\zeta$. Recall that $U$ and $D$ are the two techniquarks of the theory and $\nu'$ and $\zeta$ the extra leptons.~There is a gauge anomalous free hypercharge assignment where the charges of $U$, $D$, $\nu'$ and $\zeta$ are respectively $+1$, 0, $-1$, and $-2$. We should also emphasize that due to the fact that techniquarks transform under the adjoint representation of the gauge group, some of the Goldstone bosons are colorless di-quarks (carrying technibaryon number). Apparently $\bar{U}\bar{U}$ and $\zeta$ have charges $-2$. This candidate $HeA$ (with $A$ being $\bar{U}\bar{U}$ or $\zeta$) is a Strongly Interacting Massive Particle (SIMP), rather than a WIMP due to the large geometric cross section of the helium component. Despite the large cross section, this candidate has not been ruled out by any experiment so far. Amazingly enough, although such a candidate interacts strongly with matter, it cannot be detected in earth based detectors (based on measuring the recoil energy) like CDMS, Xenon, or LUX. By the time such a particle reaches the detector, it has lost most of its kinetic energy making it impossible to produce recoil energies above the detection threshold. In~\cite{KK2}, we examined a generalized version of the aforementioned scenario, where although the majority of dark matter is $He\bar{U}\bar{U}$ (or $He\zeta$), a small component can be of the WIMP form $\bar{\zeta}\bar{U}\bar{U}$ (or $UU\zeta$). Such a WIMP component must be small since it is constrained by direct detection experiments.

In~\cite{KK,KK2}, we had assumed that techniparticles are stable. In particular with respect to the technibaryons, the symmetry associated with the technibaryon number protected the lightest di-quark Goldstone boson from decaying. Here we reexamine the scenario of~\cite{KK2} allowing decays of the techniparticles. It has been demonstrated that decaying dark matter can provide a possible explanation of the unexpected positron excess seen in PAMELA~\cite{Nardi:2008ix,Arvanitaki:2008hq}. Decaying of dark matter particles through a dimension-6 operator gives a lifetime
\beq
\tau \sim 8 \pi \frac{M_{\text{GUT}}^4}{m^5} =5 \times 10^{20}\text{s} \left ( \frac{\text{2TeV}}{m} \right )^5 \left ( \frac{M_{\text{GUT}}}{10^{15} \text{GeV}} \right )^4,
\label{op6}
\eeq
where $m$ is the mass of the dark matter particle. Note that we have normalized the lifetime with respect to a GUT scale by an order of magnitude lower than the typical value of $2 \times 10^{16}$ GeV suggested by supersymmetry. As we are going to argue a small component of dark matter with a mass of $\sim$TeV or less and a lifetime of $10^{20}\text{s}$ can accommodate nicely the positron excess seen in PAMELA and AMS-02 data.  In addition such a lifetime is sufficiently large  in order not to deplete the density of this component of dark matter by today since it is a few orders of magnitude larger than the age of the universe. As it was stressed in~\cite{Nardi:2008ix}, dimension-6 operators are very natural objects in Technicolor, and therefore such a framework becomes very appealing.

\section{Techniparticle Excess}

We already mentioned that the MWT  has two techniquarks $U$ and $D$ in the adjoint representation of the Technicolor $SU(2)$ with charges $+1$ and 0, and  two new leptons $\nu'$ and $\zeta$ with charges $-1$ and $-2$  respectively. The theory possesses a global $SU(4)$ symmetry that breaks spontaneously to an $SO(4)$. Out of the 9 Goldstone bosons, three of them (with the quantum numbers of the usual pions) are eaten by the $W$ and $Z$ bosons, while the rest 6 are the colorless di-quarks $UU$, $UD$, and $DD$ and their antiparticles~\cite{Gudnason:2006yj}.

We are going to consider two possibilities. The first one is to have an excess of $-2$ charge $\bar{U}\bar{U}$ and a little of $+2$ $\bar{\zeta}$. The main component of dark matter is the SIMP $He\bar{U}\bar{U}$. There is also a small WIMP component of $\bar{\zeta}\bar{U}\bar{U}$. The second scenario is to have an excess of $\zeta$ and a little of $UU$, in such a way that the main SIMP component of dark matter is $He\zeta$ and the small WIMP one $UU\zeta$. In both cases we have assumed that $UU$ is the lightest among the technibaryons and similarly $\zeta$ is the lightest of the new leptons. The calculation of the relic density of the technibaryons taking into account sphaleron violating processes, weak equilibration and overall charge neutrality gives similarly to~\cite{Gudnason:2006yj}
\beq
\frac{TB}{B}=-\sigma_{UU} \left (\frac{L'}{B}\frac{1}{3\sigma_{\zeta}}+1+\frac{L}{3B} \right ),
\eeq
where $TB$, $B$, $L$, and $L'$ are the technibaryon, baryon, lepton, and new lepton family number respectively. $\sigma_i$ are statistical factors for the specific particle $i$ given by
\beq
\sigma_i = \begin{cases}
6f\left(\frac{m_i}{T^*}\right) & \textrm{for fermions} \ , \\
6g\left(\frac{m_i}{T^*}\right) & \textrm{for bosons} \ ,
\end{cases}
\eeq
where the functions $f$ and $g$ are defined as follows
\begin{align}
f(z) &= \frac{1}{4\pi^2}\int_0^\infty dx\
x^2\cosh^{-2}\left(\frac{1}{2}\sqrt{x^2+z^2}\right) \ , \label{funcf} \\
g(z) &= \frac{1}{4\pi^2}\int_0^\infty dx\
x^2\sinh^{-2}\left(\frac{1}{2}\sqrt{x^2+z^2}\right) \ . \label{funcg}
\end{align}
\begin{figure}[t]
\begin{center}
\includegraphics[width=.45\textwidth, height=0.45 \textwidth]{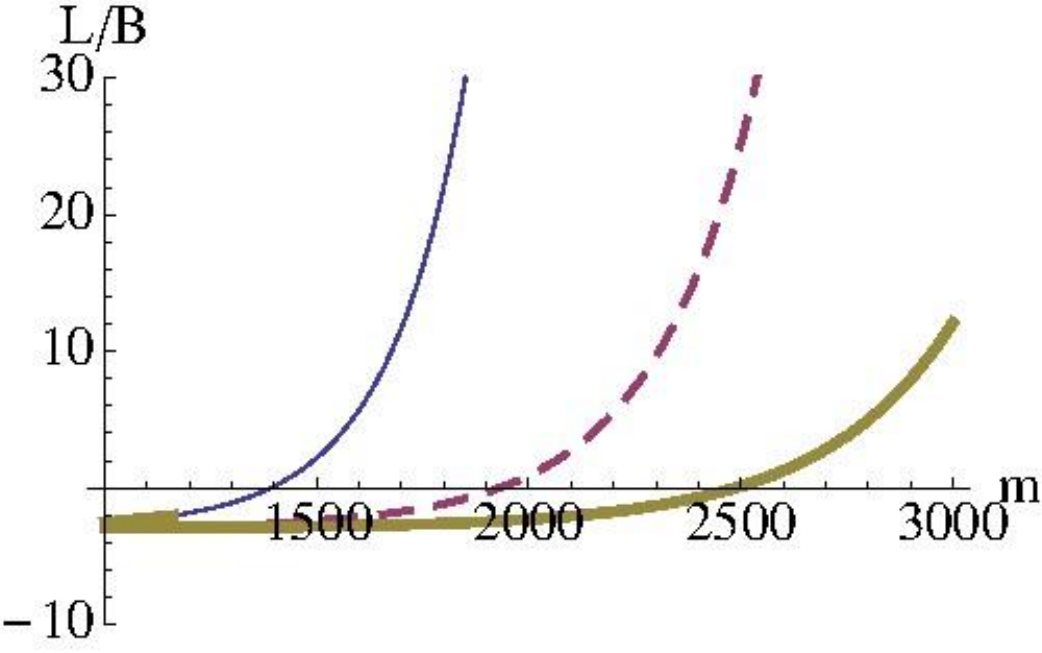}\quad \quad
%\bigskip
\includegraphics[width=.45\textwidth, height=0.45\textwidth]{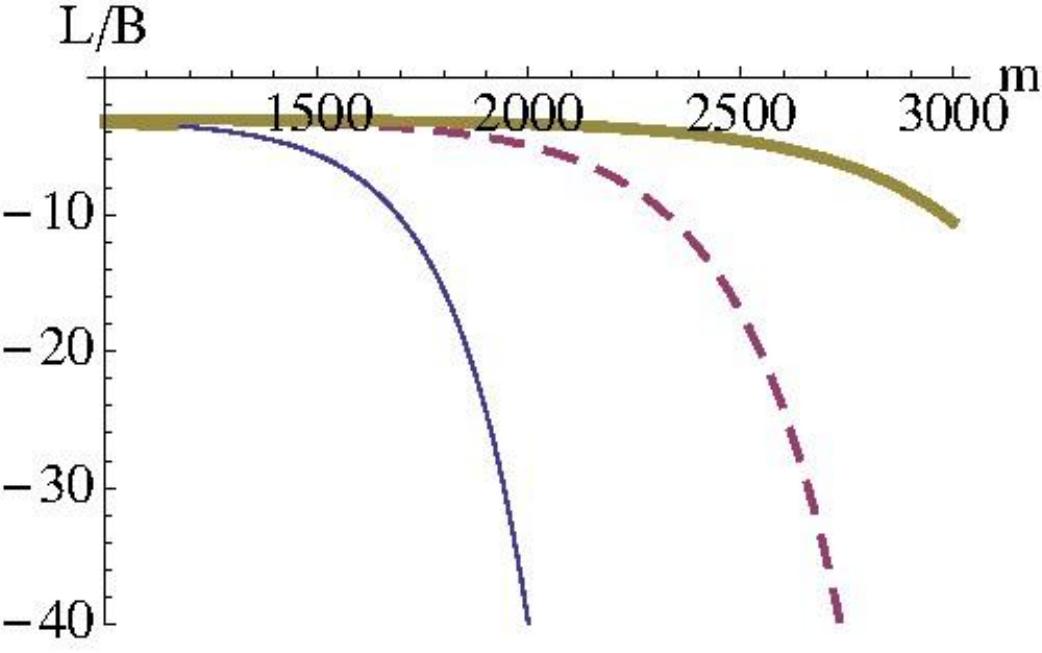}
\caption{ The ratio $L/B$ for the two scenarios explained in the text: Eqs.~(\ref{lb1}), and (\ref{lb2}) respectively for $m= m_{UU}=m_{\zeta}$ (in GeV), and $\xi=10^{-6}$.  The three different lines i.e. thin solid, dashed and thick solid correspond to freeze-out temperature for the sphalerons $T^*$ of 150, 200, and 250 GeV respectively.}
\end{center}
\end{figure}
$T^*$ is the freeze-out temperature for the sphaleron process, usually taken somewhere between 150-250 GeV.
In the first aforementioned possibility, the dark matter density is
\beq
\frac{\Omega_d}{\Omega_B}=\frac{\Omega_{\bar{\zeta}\bar{U}\bar{U}}}{\Omega_B}+
\frac{\Omega_{He\bar{U}\bar{U}}}{\Omega_B}=\left | \frac{L'}{B} \right | \frac{m_d}{m_p}+\left ( \frac{3}{2} \left |\frac{TB}{B} \right |- \left |\frac{L'}{B} \right | \right )\frac{m_s}{m_p}=5.47,
\eeq
where $m_d$, $m_s$, and $m_p$ are the masses of $UU\zeta$, $He\bar{U}\bar{U}$ and proton respectively. We have taken the ratio of dark matter to baryonic matter to be $\sim 5.47$. If $\xi$ denotes the fraction of the WIMP component ($\bar{\zeta}\bar{U}\bar{U}$) of dark matter, then the ratio of leptons over baryons is given by
\beq
\frac{L}{B}=-3 +5.47m_p \left [ \frac{\xi}{m_d \sigma_{\zeta}}+\frac{2\xi}{m_d \sigma_{UU}}+\frac{2(1-\xi)}{m_s \sigma_{UU}} \right ]. \label{lb1}
\eeq
 In the second scenario (that of $He\zeta$ and $UU\zeta$), 

\beq
\frac{\Omega_d}{\Omega_B}=\frac{\Omega_{UU \zeta}}{\Omega_B}+\frac{\Omega_{He \zeta}}{\Omega_B}=
\frac{3}{2}\frac{TB}{B}\frac{m_d}{m_p}+\left (\frac{L'}{B}-\frac{3}{2} \frac{TB}{B} \right ) \frac{m_{s'}}{m_p},
\eeq
where $m_s'$ is the mass of $He\zeta$. The ratio $L/B$ is
\beq \frac{L}{B}=-3 -5.47m_p \left ( \frac{\xi}{m_d \sigma_{\zeta}}+\frac{2\xi}{m_d \sigma_{UU}}+\frac{1-\xi}{m_{s'} \sigma_{\zeta}} \right ). \label{lb2}\eeq 
$\xi$ here is again the fraction of the WIMP-like component of dark matter.
There are two points we would like to emphasize. The first one is that both possibilities give a ratio of lepton over baryon numbers very close to $-3$ if the masses of $UU$ and $\zeta$ remain around 1 TeV. In fact the first scenario gives a ratio slightly above $-3$ and the second slightly below. $L/B$ starts deviating (exponentially) as a function of the mass of $UU$ and/or $\zeta$ once we go to masses much higher than 1.5 to 2 TeV (see Fig.~1). The second point we would like to stress is that $\xi$ is constrained by earth based direct detection search experiments. In~\cite{KK2} we found that the WIMP component of this dark matter scenario cannot be more than 1\% (or $\xi<0.01$). Since then, the constraint from the CDMS and Xenon experiments has improved significantly and more severe constraints from LUX appeared. The cross section of $UU\zeta$ (or its antiparticle) with a proton is~\cite{Goodman:1984dc}
\beq
\sigma_p=\frac{G_F^2}{2\pi}\mu^2\bar{Y}^2F^2\simeq 1.8 \times 10^{-39}\text{cm}^2,
\eeq
where $\bar{Y}=Y_L+Y_R$ i.e. the sum of the hypercharge of left and right components. It is easy to check that in our case $\bar{Y}=-1/2$. This is because $UU$ has $Y_L=1$ and $Y_R=2$ and $\zeta$ has $Y_L=-3/2$ and $Y_R=-2$. The total sum is $-1/2$. In addition since both $UU$ and $\zeta$ are much heavier than the proton, the reduced mass $\mu$ is approximately the mass of the proton. The form factor $F$ depends on the target nucleus and the recoil energy. For example for Ge detector with recoil energies between 20 to 50 keV, the form factor ranges from 0.43 to 0.72~\cite{Gudnason:2006yj}. Here in this estimate  of the WIMP-proton cross section we have set $F=1$. The results of the LUX experiment~\cite{lux} exclude WIMPs with a cross section $10^{-45}~\text{cm}^2$  for a typical WIMP mass of 1 TeV. This means that WIMPs with the cross section of $UU\zeta$ can make up only a component of $\sim10^{-6}$ or smaller of the total dark matter. Here we are going to use a typical value of $\xi=10^{-6}$.

\section{Decaying Dark Matter}

As we mentioned in the previous section, we might have a $\sim 10^{-6}$ (or less) WIMP component in our dark matter framework. This comes in the form of $\bar{\zeta}\bar{U}\bar{U}$ (first scenario) or $UU\zeta$ (second scenario). Our goal is to consider decay processes that can produce the excess of positrons seen in PAMELA and AMS-02. For this, it is generically better if the $+2$ objects decay accordingly.

In the first considered scenario we assume that  $\bar{U}\bar{U}$ is stable, and therefore the SIMP component  which consists the overwhelming part of dark matter is unaffected. On the other hand, we assume that $\bar{\zeta}^{++}$ can decay to  leptons. By construction since $\zeta$ and $\nu'$ belong to the same electroweak doublet, $\zeta$ couples to $\nu'$ and $W^-$. Since $\nu'$ is a lepton with an electric charge $-1$, it can in principle slightly mix with the usual $-1$ leptons, i.e. electrons, muons, and taus. The tiny WIMP component of dark matter made of $\bar{\zeta}\bar{U}\bar{U}$, decays due to the fact that $\bar{\zeta}$ can decay to a $W^+$ and (via $\bar{\nu}'$) to positrons, anti-muons and anti-taus. We assume that $\nu'$ is heavier than $\zeta$ and therefore the decay is suppressed. In order not to get very fast decays of $\bar{\zeta}$, the mixing of $\bar{\nu'}$ with positrons etc has to be extremely small. However this is something expected due to experimental constraints as well as due to the fact that $\nu'$ is much heavier than the leptons.  It is also expected that the mixing  between $\nu'$ and $\tau$ would be larger than $\nu'$ and $\mu$ or $\nu'$ and positrons. The decay in this scenario can be accommodated via a dimension-5 operator. However, decays of $\bar{\zeta}$ to positrons or $\mu^+ \mu^+$ or $\tau^+ \tau^+$ can lead to unwanted production of hadrons via decays of $W^+$. Therefore we focus on the second case.

In the second scenario the small WIMP component is made of $UU\zeta$. In this case we assume that $\zeta$ is stable (and no mixing with other leptons exists), but the $UU$ Goldstone boson decays via a GUT interaction. A natural dimension-6 operator that can accommodate the decay can be of the form
\beq \mathcal{O}=\frac{U^T C U \psi^T C' \psi}{\Lambda_{\text{GUT}}^2}, \eeq where $\psi$ is an electron, muon, or tau. Notice that due to the transpose instead of the bar, such an operator violates both the lepton and the technibaryon number. It allows a possible decay of $UU$ to two positrons (or two anti-muons or anti-taus\footnote{In principle we can have an even more general operator where $UU$ decays to different species of antileptons, i.e. a positron and an anti-muon etc.})
\beq UU \rightarrow e^+ + e^+. \eeq
It is understood that although $C$ and $C'$ can be generic Dirac matrices, $C$ has to be the charge conjugate matrix in order for $U^TCU$ to be the pseudo-Goldstone boson $UU$. If we require that parity is not violated by the interaction, $C'$ must also be the charge conjugate matrix. In case parity is violated, $C'$ can be $C\gamma_5$ (as it is a well known fact that $\psi^T C\gamma_5 \psi$ is a scalar). Of course nothing forbids a similar decay of $UU$ to two quarks or even a quark and a lepton, as it would depend on the details of the GUT interaction. However here we do not want to speculate regarding the GUT interactions, but simply to demonstrate that such a realization can in fact produce the positron spectrum seen by experiments. As we already mentioned, a dimension-6 operator of the above form would give according to Eq.~(\ref{op6}) a lifetime of the order of $10^{20}$ $s$ for a mass of $UU$ of the order of TeV. If $UU$ does not decay to hadrons, this scenario is more appropriate for explaining the positron excess compared to the first scenario we mentioned because in the first scenario the decay of $\bar{\zeta}$ will always be accompanied by hadronic decays that are not seen by PAMELA.

\section{Positron Excess and Fit to the PAMELA and AMS-02 data}
Here we show the impact of decaying $UU$  particles on the cosmic positron flux and diffuse gamma radiation. The so-called ``PAMELA anomaly" in the cosmic positron spectrum~\cite{PAMELA} has been recently confirmed also by AMS-02 \cite{AMS-2}. This anomaly cannot be explained by positrons of only secondary origin, and therefore primary positron sources are needed to explain the data. There are attempts to realize it based on decaying or annihilating dark matter models. Any scenario that provides positron excess is constrained by other observational data, mainly from the data on cosmic antiprotons, and gamma-radiation from our halo (diffuse gamma-background) and other galaxies and clusters \cite{constraint1,constraint2,constraint3,constraint4,constraint5,constraint6,constraint7,Silk}. 
If dark matter does not produce antiprotons, then the diffuse gamma-ray background gives the most stringent and model-independent constraints.

\begin{figure}
\includegraphics[scale=0.7]{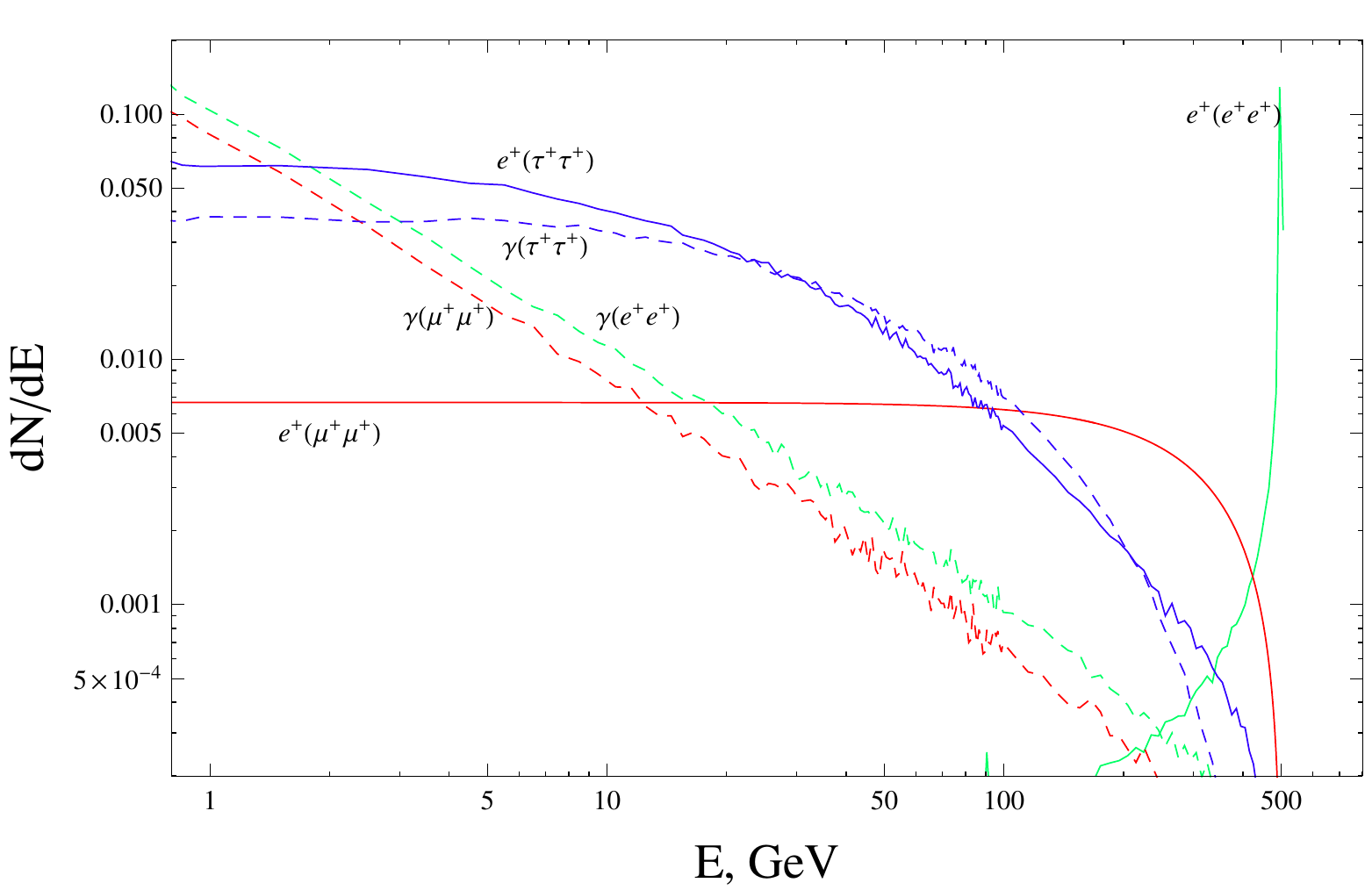}
\caption{Spectra of gamma-rays and positrons from decays $UU\rightarrow e^+e^+, \mu^+\mu^+, \tau^{+}\tau^{+}$. We used Pythia 6.4 \cite{pythia}.}

\label{init}
\end{figure}

In our scenario the $UU$ component of a tiny $UU\zeta$ WIMP component of dark matter decays as $UU\rightarrow e^+e^+, \mu^+\mu^+, \tau^+\tau^+$ in principle with different branching ratios. All decay modes give directly or cascading positrons and gamma photons, which are hereafter referred  as final state radiation (FSR). 
In Fig.~\ref{init} we show the produced positron and gamma spectra for each decay mode individually.
 Note that unlike PAMELA, the AMS-02 disfavors decays purely to $e^+$ and $\mu^+$ (although does not exclude them).

In the context of indirect dark matter searches from cosmic rays (CR), the %$\tau^+$ 
leptonic decay modes have been studied extensively (see e.g. references \cite{constraint1,constraint2,constraint3,constraint4,constraint5,constraint6,constraint7,Silk}), using a variety of different approaches in estimating the CR signals. For our estimate, we adopt the following model of positron  propagation  in the Galaxy. Due to energy losses, positrons have a finite diffusion length at given energy $E$
\begin{equation}
\lambda\sim \sqrt{\int Ddt}=\sqrt{\int D\frac{dE}{b}}\sim 10\, {\rm kpc}\sqrt{E^{-0.7}-E_0^{-0.7}},
\label{lambda}
\end{equation}
where $D\approx 4\cdot 10^{28}\, {\rm cm^2s^{-1}}E^{0.3}$ is a typical value for the diffusion  coefficient \cite{Moskalenko}, $b=\beta E^2$ is the rate of energy losses with $\beta \sim 10^{-16}$ s$^{-1}$GeV$^{-1}$, and $E_0$ is the initial energy. All energies are measured in GeV. The effect of the diffusion in the propagation can be estimated by assuming a homogeneous distribution of the sources. In fact, the result of diffusion is not sensitive to the effects of inhomogeities, because it depends on the averaged density within the diffusion length. Since we are interested in positron energies above $\sim 10$ GeV, which corresponds to $\lambda\lesssim 5$ kpc (see Eq.\eqref{lambda}) over which no essential inhomogeneity effects are expected, this simple approximation we make here is good. At $E\lesssim 10$ GeV secondary positrons dominate the spectrum. If $\lambda$ exceeds the size of the magnetic halo (MH) ($h\sim 4$ kpc in height, and $R\sim 15$ kpc in width) the leakage of particles from the halo should be taken into account. We consider this effect by introducing a suppression factor, which is equal to the ratio of the volume of MH contained  within the sphere of radius $\lambda$:
\begin{equation}
Q=1-\frac{(\lambda-h)^2(2\lambda+4)}{2\lambda^3}\eta(\lambda-h)-\frac{2h(\lambda^2-r^2)}{3\lambda^3}\eta(\lambda-R),
\end{equation}
where $\eta$ is the step function.
If $dN/dE_0$ is the number of positrons produced in a single decay (see Fig.\ref{init}), the positron flux near the Earth can be estimated as
\begin{equation}
F(E)=\frac{c}{4\pi}\frac{n_{\rm loc}}{\tau}\frac{1}{\beta E^2}
\int_E^{m/2} \frac{dN}{dE_0} Q(\lambda(E_0,E))dE_0,
\end{equation}
where $n_{\rm loc}=\xi\cdot (0.3\,{\rm  GeV/cm^3})m_{UU}^{-1}$ is the local number density of $UU$ particles with $\xi=10^{-6}\xi_{-6}$. Recall that $\xi$ is the fraction of dark matter in the WIMP $UU\zeta$ component.

\begin{figure}
\includegraphics[scale=0.7]{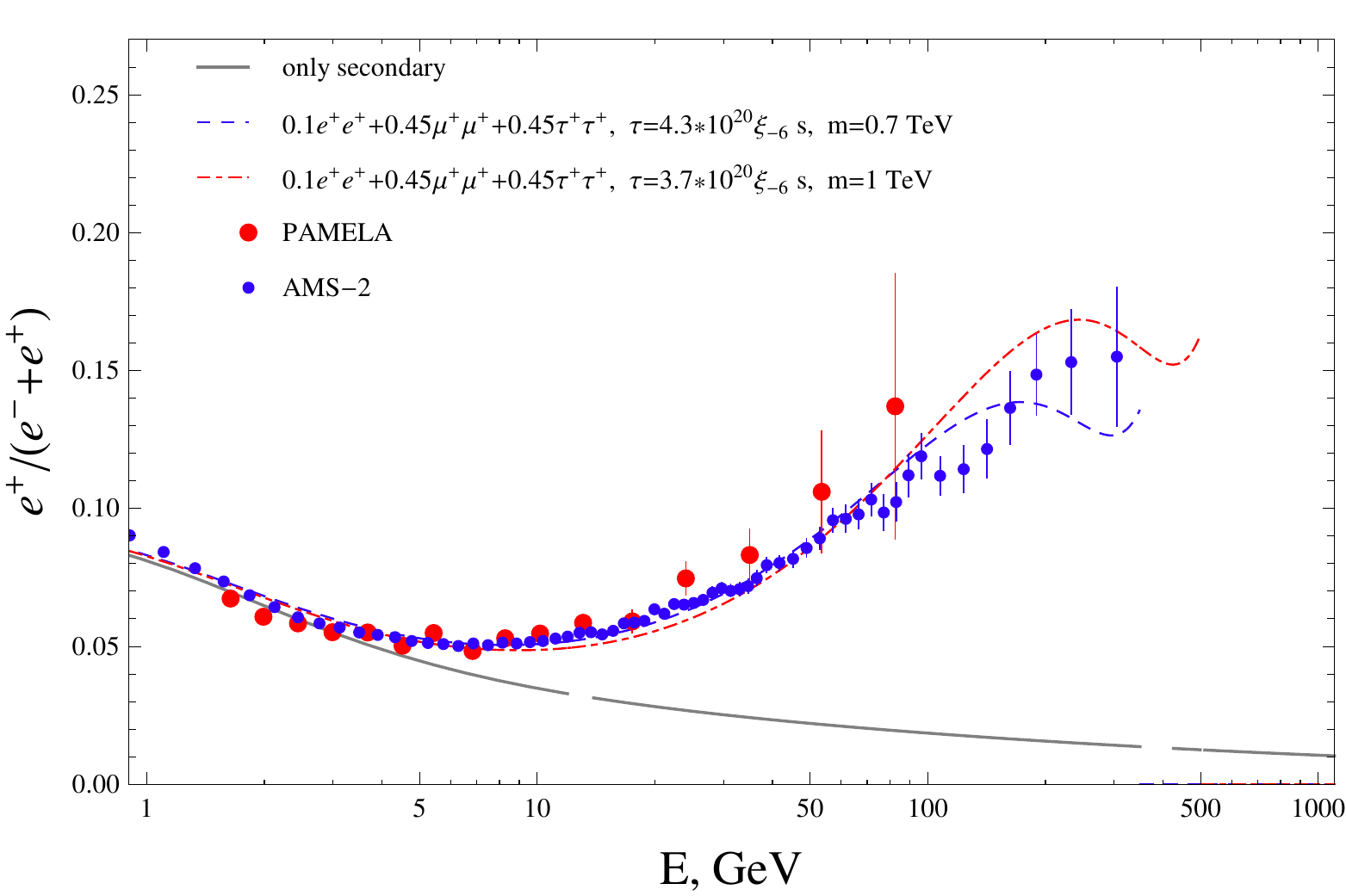}
\caption{Positron excess due to $UU\rightarrow e^+e^+, \mu^+\mu^+, \tau^+ \tau^+$ decays compared to PAMELA and AMS-02 data.}
\label{positrons}
\end{figure}

The effect of solar modulation becomes important at the less interesting low energy part of the positron spectrum. To account for this effect, we have  adopted the forced field model \cite{Gleeson} with two different $\phi$ parameters for positrons and electrons. They are easily adjusted so they can fit the data points at low energy. The positron and electron background components  were taken from \cite{Baltz}.  In Fig.~\ref{positrons} we present the positron excess due to $UU$ decays for two values of the mass of $UU$, $m_{UU}=0.7$ TeV and $m_{UU}=1$ TeV. We also show the lifetime  of $UU$ $\tau$ and the branching ratios that fit the experimental data optimally for each choice of $m_{UU}$. They evade the existing constraints of \cite{constraint1,constraint2,constraint3,constraint4,constraint5,constraint6,constraint7,Silk}.

The gamma-ray flux from $UU$ decays has two main contributions: one from FSR (shown in Fig.~\ref{init}) and another one from inverse Compton (IC) scattering of positrons  on background photons (star light, infrared background, CMB).

\begin{figure}
\includegraphics[scale=0.7]{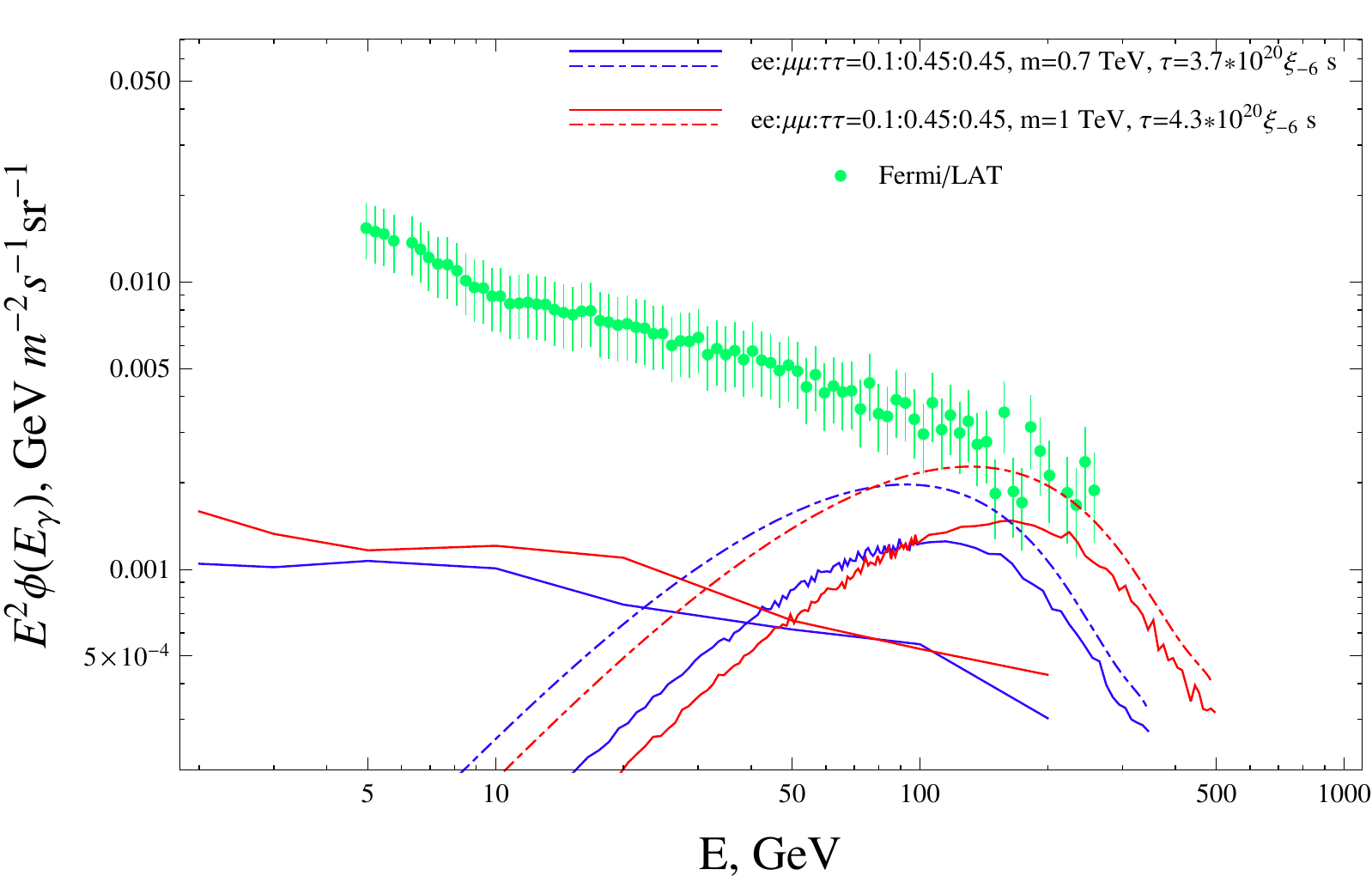}
\caption{Gamma-ray flux from $UU$ decays in the Galaxy ($|b|\ge 10^{0}$) in comparison to the Fermi/LAT data on diffuse background \cite{Fermi}. Two contributions are shown:  IC (left curves) and FSR (right curves). Dot-dashed curves takes into account FSR photons of both galactic and extragalactic origin.}

\label{gamma}
\end{figure}

For the FSR photons produced by $UU$ decays in our Galaxy, the flux arriving in the Earth is given by
\begin{equation}
F_{\rm FSR}=\frac{n_{\rm loc}}{\tau}\frac{1}{4\pi \Delta\Omega_{\rm obs}}\int_{\Delta\Omega_{\rm obs}}\frac{n(r)}{n_{\rm loc}}dld\Omega \cdot \frac{dN_{\gamma}}{dE},
\end{equation}
where we use an  isothermal profile  $\frac{n(r)}{n_{\rm loc}}=\frac{(5{\rm \,kpc})^2+(8.5{\rm \,kpc})^2}{(5{\rm \,kpc})^2+r^2}$, $r$ and $l$ are the distances from the Galactic center and the Earth respectively. We obtain the averaged flux over the solid angle $\Delta\Omega_{\rm obs}$ corresponding to $|b|>10^\circ$, $0<l<360^{\circ}$. For the IC photons from our Galaxy, we have estimated the contribution following \cite{Cirelli}.
In Fig.~\ref{gamma} we show both contributions in the gamma-ray flux for the same parameters as in Fig.~\ref{positrons}. 

Out of our Galaxy decays of $UU$ homogeneously distributed over the Universe should also contribute to the observed  gamma-ray flux. For FSR photons this contribution can be estimated as
\begin{eqnarray}
F_{\rm FSR}^{\rm (U)}(E)=\frac{c}{4\pi}\frac{\langle n_{\rm mod}\rangle}{\tau}\int \frac{dN}{dE}dt= 
\frac{c\langle n_{\rm mod}\rangle}{4\pi\tau}\times\nonumber\\ \times \int_0^{\min(1100,\frac{m}{2E}+1)} \frac{dN}{dE_0}(E_0=E(z+1))\frac{H_{\rm mod}^{-1}dz}{\sqrt{\Omega_{\Lambda}+\Omega_m(z+1)^3}},
%\frac{3\langle n_{\rm mod}\rangle ct_{\rm mod}h}{8\pi\tau}\times\nonumber\\ \times \int_0^{\min(1100,\frac{m}{2E}+1)} \frac{dN}{dE_0}(E_0=E(z+1))\frac{dz}{\sqrt{\Omega_{\Lambda}+\Omega_m(z+1)^3}},
\label{FU}
\end{eqnarray}
where $z=1100$ corresponds to the recombination epoch, $\langle n_{\rm mod}\rangle$ is the current cosmological number density of $UU$, $H_{\rm mod}^{-1}=\frac{3}{2}t_{\rm mod}\sqrt{\Omega_{\Lambda}}\ln\left(\frac{1+\sqrt{\Omega_{\Lambda}}}{\sqrt{\Omega_{m}}}\right)$ is the inverse value of the  Hubble parameter with $t_{\rm mod}$ being the  age of the Universe. %$h=\sqrt{\Omega_{\Lambda}}\ln\left(\frac{1+\sqrt{\Omega_{\Lambda}}}{\sqrt{\Omega_{m}}}\right)$, 
$\Omega_{\Lambda}$ and $\Omega_m=1-\Omega_{\Lambda}$ are respectively the current vacuum and matter relative densities. Note in Eq.(\ref{FU}) the transition between distributions at different $z$, $ \frac{dN}{dE}\rightarrow  \frac{dN}{dE_0}(z+1)$. This extragalactic contribution to FSR increases significanlty the  total gamma-ray flux  as shown in  Fig.\ref{gamma} by dot-dashed lines.

It is not expected that extragalactic IC photons can contribute significantly to the spectrum. Indeed, mainly only low energetic CMB photons are present in the medium outside the Galaxy (or before the galactic stage). After the scattering of electrons with energy $E_0\lesssim 500$ GeV off CMB photons with energy $\omega_{\rm CMB}\lesssim 10^{-3}(z+1)$ eV,   the recoiled photons acquire at  redshift $z$  energy $\omega\sim (E_0/m_e)^2\omega_{\rm CMB}\lesssim (z+1)$ GeV, which is below 1 GeV in the modern epoch.
%where ``$z+1$" is eaten back to the modern epoch. 
It makes therefore this contribution indifferent for the energy range of Fermi/LAT.

Concluding, on the basis of Fig.\ref{gamma} one may assert that the considered scenarios of $UU$ decays satisfy the Fermi/LAT constraints. In addition, although we used the best fit values for the branching ratios, we have found that some small variation of the branching ratios is possible. If one chooses $m_{UU}>1$ TeV a possible satisfaction of the constraints is possible at the expense of the positron spectrum fit.

\section{Conclusions}

%\medskip
Dark matter can potentially be in the form of neutral $OHe$ dark atoms made of stable heavy doubly charged particles and primordial He nuclei bound by ordinary Coulomb interactions. This scenario sheds new light on the nature of dark matter and offers a nontrivial solution for the puzzles of direct dark matter searches. It can be realized in the framework of Minimal Walking Technicolor, in which an exact relation between the dark matter density and baryon asymmetry can be naturally obtained predicting also the ratio of leptons over baryons in the Universe.
 In the context of this scenario a sparse component of WIMP-like dark atoms of charged techniparticles can also appear. Direct searches for WIMPs put severe constraints on the presence of this component. However we demonstrated in this paper that
 the existence of a metastable positively doubly charged techniparticle, forming this tiny subdominant WIMP-like dark atom component and satisfying the direct WIMP searches constraints, can play an important role in the indirect effects of dark matter. We found that decays of such positively charged constituents of WIMP-like dark atoms to the leptons $e^+e^+, \mu^+\mu^+,\tau^+ \tau^+$ can explain the observed excess of high energy cosmic ray positrons, while being compatible with the observed gamma ray background. These decays are naturally  facilitated by GUT scale interactions. This scenario makes a prediction about the ratio of leptons over baryons in the Universe to be close to $-3$. The best fit of the data takes place for a mass of this doubly charged particle of 1 TeV or below making it accessible in the next run of LHC.

 {\bf Acknowledgements} C.K. is supported by the Danish National Research Foundation, Grant No. DNRF90.

\end{document}